\PassOptionsToPackage{svgnames, prologue, dvipsnames}{xcolor}

\documentclass{egpubl}
\usepackage{mlvis2026}

\WsPaper      

\accepted

\usepackage[T1]{fontenc}
\usepackage{dfadobe}

\biberVersion
\BibtexOrBiblatex
\usepackage[backend=biber,bibstyle=EG,citestyle=alphabetic,backref=true]{biblatex} 
\electronicVersion
\PrintedOrElectronic
\ifpdf \usepackage[pdftex]{graphicx} \pdfcompresslevel=9
\else \usepackage[dvips]{graphicx} \fi

\usepackage{egweblnk}

\usepackage{tabu}                      
\usepackage{booktabs}                  
\usepackage{lipsum}                    
\usepackage{mwe}            

\usepackage{csquotes}
\usepackage{amsmath}
\usepackage{bm}

\usepackage{cleveref}
\usepackage{fontawesome}
\usepackage[nohypertypes={acronym}]{glossaries}
\usepackage{xspace}
\usepackage{multirow}
\usepackage{mathtools} 
\usepackage{wrapfig}

\usepackage{mathptmx}                  
\usepackage{rotating}

\usepackage{pgfplots}    

\usepackage{hyperref}

\usepackage{xcolor}
\DeclareUnicodeCharacter{2212}{−}

\glsdisablehyper

\usepgfplotslibrary{groupplots,dateplot}
\pgfplotsset{compat=1.18}
\usetikzlibrary{shapes,shapes.misc, graphs, patterns,shapes.arrows, arrows.meta, positioning, fit, arrows.meta, graphs, shapes.misc, matrix, quotes, patterns, shapes.arrows, shapes.misc, positioning, calc, fit, bending}
\graphicspath{{figures/}{pictures/}{images/}{./}} 

\newcommand{\X}{\ensuremath{\bm{X}}}
\newcommand{\Y}{\ensuremath{\bm{Y}}}
\newcommand{\Z}{\ensuremath{\bm{Z}}}

\newcommand{\x}{\ensuremath{\bm{x}}}
\newcommand{\y}{\ensuremath{\bm{y}}}
\newcommand{\z}{\ensuremath{\bm{z}}}

\newcommand{\sZ}{\ensuremath{\bm{Z}_s}}

\newcommand{\abs}[1]{\ensuremath{\left\lvert #1 \right\rvert}}
\newcommand{\norm}[1]{\ensuremath{\left\lVert #1 \right\rVert}}

\newcommand{\demolink}{https://hereditary.cgv.tugraz.at/param-inter}
\newcommand{\codelink}{https://github.com/Dakantz/ParamInter}

\definecolor{tsnebg}{RGB}{250, 254, 254}

\definecolor{SelInput}{named}{Moccasin}
\definecolor{SelOutput}{named}{LightCoral}

\newenvironment{egauthorlist}{%
  \setlength\topsep{0pt}
  \setlength\parskip{0pt}
  \begin{minipage}{\textwidth}
    \begin{center}
      }{
    \end{center}
  \end{minipage}
}

\newcounter{@affiliationcounter}
\newcommand{\atpa}[1]{%
  \ifcsname the@affil#1\endcsname
  \else
    \ifcsname @icmlsymbol#1\endcsname
    \else
      \stepcounter{@affiliationcounter}%
      \newcounter{@affil#1}%
      \setcounter{@affil#1}{\value{@affiliationcounter}}%
    \fi
  \fi%
  \ifcsname @icmlsymbol#1\endcsname
    \textsuperscript{\csname @icmlsymbol#1\endcsname\,}%
  \else
    \textsuperscript{\arabic{@affil#1}\,}%
  \fi
}

\newcommand{\EGauthor}[3]{%
  \ifdefined\isaccepted
    #1\,\foreach\theaffil in {#2} {\atpa{\theaffil}}\ifthenelse{\equal{#3}{}}{}{\orcid{#3}}%
  \else

    \ifdefined\anonauthprinted
      \typeout{Printed anon already!}%
    \else
      \mbox{Anonymous Authors}\atpa{anon}
      \newcommand{\anonauthprinted}{def}
    \fi
  \fi
}
\newcommand{\EGaffiliation}[2]{%
  \ifdefined\isaccepted
    \ifcsname the@affil#1\endcsname
      \\ \csname @affilname \endcsname \csname the@affil#1 \endcsname \textsuperscript{\arabic{@affil#1}}{#2}%
    \else
      {\bf AUTHORERR: Error in use of \textbackslash{}icmlaffiliation command. Label ``#1'' not mentioned in some \textbackslash{}icmlauthor\{author name\}\{labels here\} command beforehand. }
      \typeout{}%
      \typeout{}%
      \typeout{*******************************************************}%
      \typeout{Affiliation label undefined. }%
      \typeout{Make sure \string\icmlaffiliation\space follows }
      \typeout{all of \string\icmlauthor\space commands}%
      \typeout{*******************************************************}%
      \typeout{}%
      \typeout{}%
    \fi
  \else 
    \ifdefined\anonprinted
      \typeout{Printed anon already!}%
    \else
      \\ \textsuperscript{1}{Anonymous Institution, Anonymous Country}%
      \newcommand{\anonprinted}{def}
    \fi %
  \fi
}

\newacronym[shortplural=LMs, longplural=Language Models]{lm}{LM}{Language Model}
\newacronym[shortplural=KGs, longplural=Knowledge Graphs]{kg}{KG}{Knowledge Graph}
\newacronym[shortplural=GUIs, longplural=Graphical User Interfaces]{gui}{GUI}{Graphical User Interface}
\newacronym{ged}{GED}{Graph Edit Distance}
\newacronym{rag}{RAG}{Retrieval Augmented Generation}
\newacronym{oql}{OQL}{Ontology Query Language}
\newacronym[shortplural=BGPs, longplural=Basic Graph Patterns]{bgp}{BGP}{Basic Graph Pattern}
\newacronym{gbnf}{GBNF}{GGML Backus-Naur Form}
\newacronym{onset}{OnSET}{Ontology and Semantic Exploration Toolkit}
\newacronym{bto}{BTO}{Brainteaser Ontology}
\newacronym{als}{ALS}{Amyotrophic lateral sclerosis}
\newacronym{ms}{MS}{Multiple sclerosis}
\newacronym{pca}{PCA}{Principal Component Analysis}

\newacronym[shortplural=IAIs, longplural=interpretable artifical intelligence]{iai}{IAI}{interpretable artificial intelligence}
\newacronym[shortplural=DNNs, longplural=deep neural networks]{dnn}{DNN}{deep neural network}
\newacronym[shortplural=NNs, longplural=neural networks]{nn}{NN}{neural network}
\newacronym[shortplural=EAFs, longplural=electric arc furnaces]{eaf}{EAF}{electric arc furnace}
\newacronym[shortplural=ALEs, longplural=averaged local effects]{ale}{ALE}{averaged local effects}
\newacronym[shortplural=PDPs, longplural=partial dependence plots]{pdp}{PDP}{partial dependence plot}
\newacronym{ml}{ML}{Machine Learning}
\newacronym{ai}{AI}{Artificial Intelligence}
\newacronym{lime}{LIME}{local interpretable model-agnostic explanations}
\newacronym{knn}{kNN}{$k$-Nearest Neighbors}
\newacronym{shap}{SHAP}{Shapley additive explanations}
\newacronym{som}{SOM}{Self-Organizing Maps}
\newacronym{iid}{iid.}{independent and identically distributed}
\newacronym{mse}{MSE}{mean squared error}
\newacronym{mae}{MAE}{mean absolute error}
\newacronym{relu}{ReLU}{rectified linear unit}
\newacronym{xai}{XAI}{eXplainable Artificial Intelligence}
\newacronym{sg}{SG}{SmoothGrad}
\newacronym{ig}{IG}{integrated gradients}
\newacronym{ofc}{OFC}{optical flow constraint}
\newacronym[shortplural=GMMs, longplural=Gaussian mixture models]{gmm}{GMM}{Gaussian mixture model}
\newacronym[shortplural=GPs, longplural=Gaussian processes]{gp}{GP}{Gaussian process}
\newacronym{ood}{OOD}{out-of-domain}
\newacronym{ucq}{UCQ}{Uncertainty Quantification}
\newacronym[shortplural=LGBMs, longplural=Light Gradient Boosting Machines]{lgbm}{LightGBM}{Light Gradient Boosting Machine}
\newacronym{map}{MAP}{maximum a posteriori}
\newacronym{kl}{KL}{Kullback-Leibler}
\newacronym{mc}{MC}{monte-carlo}
\newacronym{gda}{GDA}{gaussian discriminatory analysis}
\newacronym{ddu}{DDU}{deep deterministic uncertainty}
\newacronym[shortplural=DEs, longplural=deep ensembles]{de}{DE}{deep ensemble}
\newacronym{clue}{CLUE}{counterfactual latent uncertainty explanations}
\newacronym{vae}{VAE}{Variational Autoencoder}
\newacronym{rbf}{RBF}{radial basis function}
\newacronym{se}{SE}{squared exponential}
\newacronym{tsne}{t-SNE}{t-Distributed Stochastic Neighbor Embedding}
\newacronym[shortplural=SCAs, longplural=smoothness constrained attributions]{sca}{SCA}{smoothness constrained attribution}
\newacronym[shortplural=PIs, longplural=prediction intervals]{pi}{PI}{prediction interval}
\newacronym[shortplural=PIOs, longplural=prediction interval overlap]{pio}{PIO}{prediction interval overlap}
\newacronym[shortplural=PICPs, longplural=prediction interval coverage probability]{picp}{PICP}{prediction interval coverage probability}
\newacronym[shortplural=iUCAMs, longplural=input uncertainty attribution mechanisms]{iuam}{iUCAM}{input uncertainty attribution mechanism}
\newacronym{mast}{MAST}{Multi-Agent Spatio-Temporal}
\newacronym[shortplural=pdfs, longplural=probability density functions]{pdf}{pdf}{probability density function}
\newacronym[shortplural=CIs, longplural=Confidence Intervals]{ci}{CI}{Confidence Interval}

\glsenableentrycount

\newcommand{\toolname}{ParamInter\xspace}
\title[Parameter Space Analysis through Guided Visual Interpolations]{Parameter Space Analysis through Guided Visual Interpolations}
\author[Kantz et.al ]
{
\begin{egauthorlist}
    \EGauthor{Benedikt Kantz}{tug}{0000-0003-3294-8421}
    ,
    \EGauthor{Peter Waldert}{tug}{0009-0004-8459-7381},
    \EGauthor{Stefan Lengauer}{tug}{0000-0001-5136-4320}
    \EGauthor{Clemens Staudinger}{voest}{},
    \EGauthor{Stefan Schuster}{voest}{},
    and 
    \EGauthor{Tobias Schreck}{tug}{0000-0003-0778-8665}
\end{egauthorlist}
        \\
    \EGaffiliation{tug}{Graz University of Technology, Austria}
    \EGaffiliation{voest}{voestalpine Stahl GmbH, Linz, Austria}
}
\begin{document}

\teaser{
  \centering
  \pgfdeclarelayer{background}
\pgfdeclarelayer{foreground}
\pgfsetlayers{background,main,foreground}
\usepgfplotslibrary{groupplots,dateplot}
\usetikzlibrary{shapes,shapes.misc, graphs, patterns,shapes.arrows, arrows.meta, positioning, fit, arrows.meta, graphs, shapes.misc, matrix, quotes, patterns, shapes.arrows, shapes.misc, positioning, calc, fit, bending}

\newcommand{\boxheight}{2.4cm}
\newcommand{\boxwidth}{2.6cm}
\begin{tikzpicture}[
  block/.style={rectangle, draw=LightCoral, fill=LightCoral!5, minimum size=1cm,  text width=\boxwidth, align=center, font={\small},  line width=1.5pt, minimum height=2cm,
      inner sep=2mm, },
  >={Latex[round]},
  every new ->/.style={shorten >=1pt, line width=1pt},
  every new --/.style={line width=1pt},
  ]
  \node(input_sel)[block] {$a.$ Input selection ($\x_0$) \\\includegraphics[height=\boxheight]{nobg/blast_furnace/01a_selection.png}};
  \coordinate (input_sel_bel) at ([yshift=-0.5cm]input_sel.south);
  \node(output_sel)[block, draw=Moccasin, fill=Moccasin!5, right= 2cm of input_sel,  text width=5.5cm]{\centering  $b.$  Guided cost function adjustments ($o(\z)$)\\ \includegraphics[height=\boxheight]{interface/02_guidance.pdf}\includegraphics[height=\boxheight]{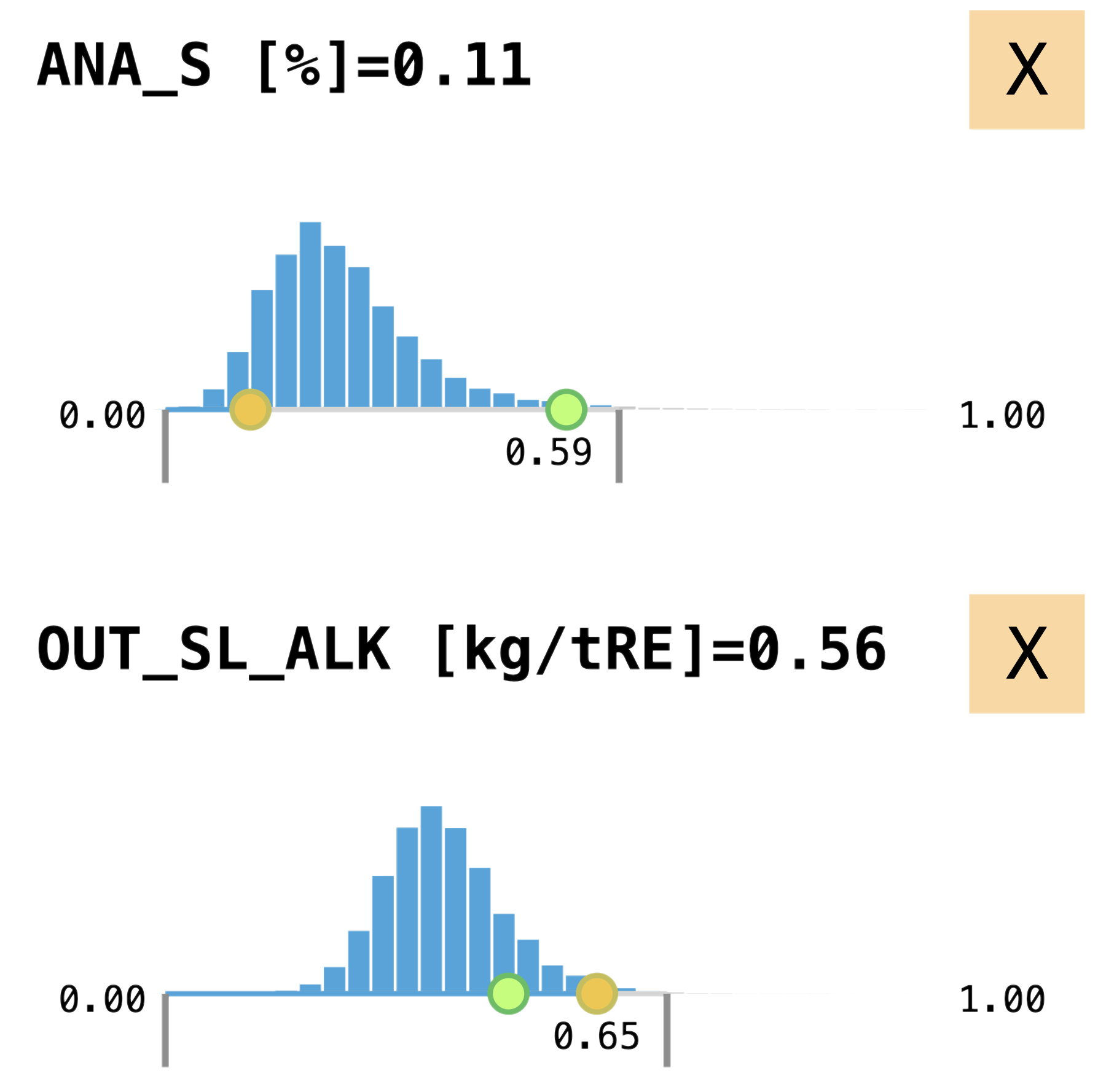}};
  \node(interpolation_result)[block, right=2cm of output_sel, draw=SkyBlue, fill=tsnebg, text width=4.7cm]{\centering $c.$ Interpolation Overview ($\z_{\lambda}$)\\ \includegraphics[height=\boxheight]{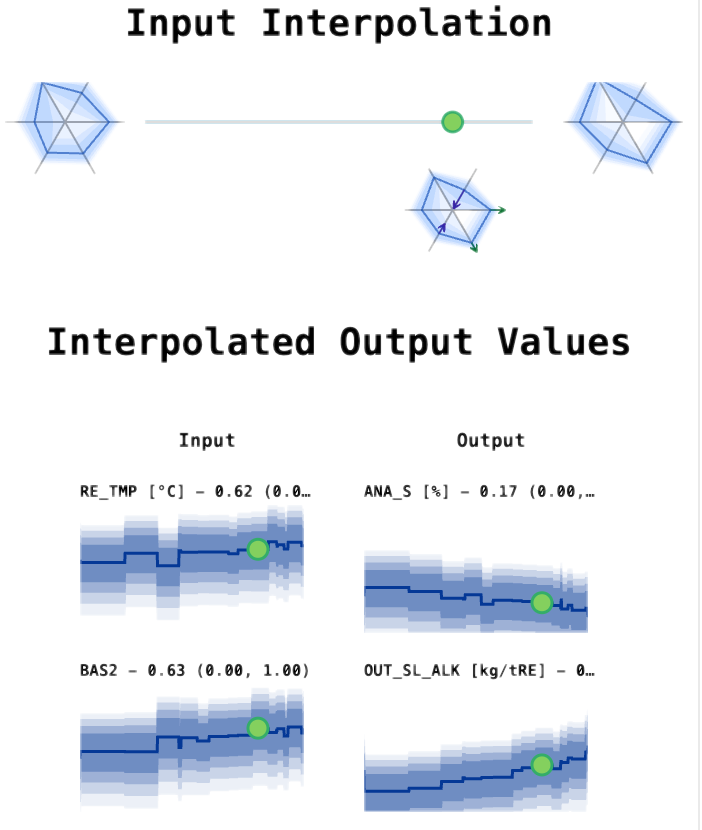} \includegraphics[height=\boxheight]{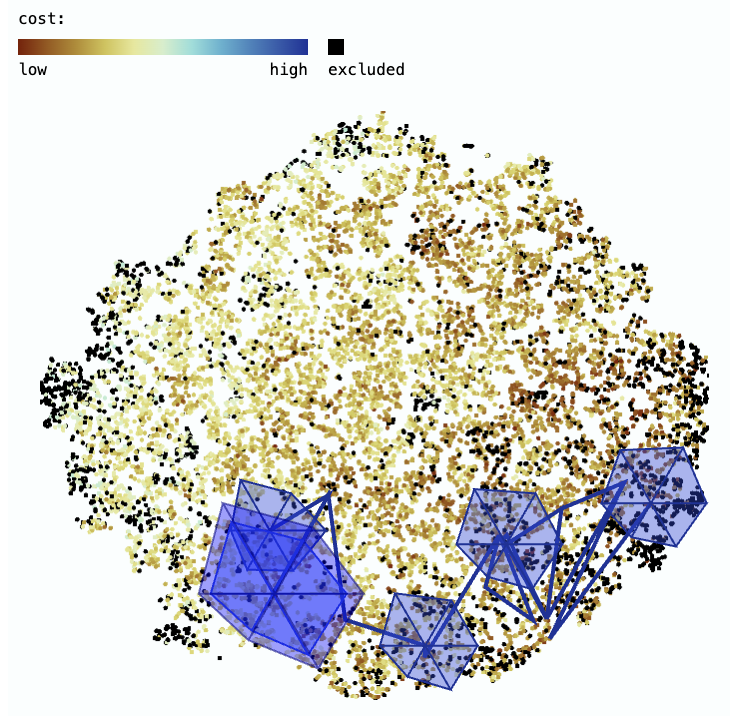}};
  \coordinate (interpolation_result_bel) at (input_sel_bel-|interpolation_result.south);
  \graph [use existing nodes, edge quotes={font={\small}, anchor=south, align=center, text=black}] {
  input_sel->["similarity\\ search"]output_sel->["minimize $o(\z)$ \& \\ interpolate"]interpolation_result;
  interpolation_result--interpolation_result_bel--["refine cost, update start"]input_sel_bel->input_sel;
  };
\end{tikzpicture}

  \caption{Our proposed exploration workflow within the \toolname tool. The input selection $(a.)$  allows the users to select their initial parameters. Next, the guided cost function adjustments $(b.)$ foster the understanding of the dependencies between input parameters and output parameters, while allowing data selection and setup of a cost function. Once the intended cost function has been set up, the interpolation overview ($c.$) shows a possible path between the start and intended end composition in the data and embedding space. The colors  of the embedding space represent the value of the cost function, showing regions with low costs directly.}
  \label{fig:teaser}
}

\maketitle
\begin{abstract}
  We propose Parameter Space Analysis through Guided Visual Interpolations (\toolname), a novel tool for high dimensional input parameter space analysis by making interpolation towards optimal parameter sets explorable using guided analytics. The interpolation is accompanied by both small multiples in linked views, and utilizes \cgls{tsne} representations to show an interpolation overview. \toolname uses a guided exploration loop focusing on the interpolation towards user-specified target parameters from many output parameters.
  The exploration process is additionally guided through \cgls{xai}-based effect suggestions throughout our tool.
  \toolname, compared to prior work, focuses on the integration of state-of-the art effect-based \cgls{xai} and \cgls{ucq} approaches for guidance,  and introduces an interpolation towards the optimal solution through interpolation between the initial parameter setting and the optimal setting. We also add an interpretability layer for dimensionality-reduced data by displaying our novel interpolation towards the optimum, enhanced by small multiples of the input parameters on top.
  We demonstrate the direct applicability of our tool on a real-world use case for a blast furnace optimisation process, where a multi-objective problem is solved through modeling and visualisation. 

\end{abstract}

\section{Introduction}

The discovery of optimal or advantageous data points within high-volume, high-dimensional datasets can be formulated either mathematically as minimisation problems or via iterative filtering. Our problem setting is motivated by a typical input-output model, where continuous input parameters $\x\in\X$ drive an unknown process with multiple outputs $\mathbf{f}(\x)$, yielding continuous $\y\in\Y$ output vectors, i.e. a basic input-output model~\cite{Sedlmair2014VisualParameterSpaceAnalysis}. These can be optimised under a specific cost function $o(\y)$, i.e. $\arg\min_{\x} o(\mathbf{f}(\x))$ to obtain the best parameter set. Approaches within visualisation, and visual analytics specifically, often employ parallel coordinates or turn towards the Pareto front~\cite{Cajot2019InteractiveOptimizationParallelCoordinates,Yang2003InteractiveHierarchicalDisplays,Cibulski2020ParetoFrontVisualization,Chen2013SelfOrganizingPareto}. 
We, however, turn in our visual parameter space analysis towards optimisation of the parameter through interpolations, as well as uncertainties and sensitivity analysis within the space~\cite{Sedlmair2014VisualParameterSpaceAnalysis}. 
Our tool, \toolname, through these aspects, fosters the understanding of the progression towards the optimal solution by providing an interface to smoothly interpolate between a chosen start configuration and an optimisation target. We additionally employ \cgls{xai} as a guidance approach to provide a local hint of relevant input parameters interactively the cost function composition. \toolname additionally, integrates \cgls{ucq} into the visualisations, providing an indication of the confidence within the data itself, and supporting the decision-making process~\cite{Sarma2024OddsInsightsUCQ,Marusich2024UsingAIUncertaintyQuantification}.

Our approach centers on the use of glyphs, namely radar charts, as a mediator for setting the initial input parameter search and for displaying input parameter sets. We also utilize the space within the glyphs for our \cgls{xai} and \cgls{ucq} displays whenever appropriate, guiding the users in their analysis. The radar charts also show the interpolation while the user scrubs through it interactively, strengthening the interpretation of the interpolation. The glyphs are supported by small multiples of histograms for brushing and composition of the optimisation function, and line charts with uncertainty displays. Our interpolation is, finally, also drawn over \cgls{tsne} plots, as shown in \Cref{fig:teaser} $c.$. Each \cgls{tsne} plot is concerned with one input- or output group, to aid with a clearer understanding of the relations. The interpolation and visualisations update on each change to the cost function $o(\z$) and filter $s(\Z)$, enhancing understanding of the data composition and optimal parameter regions, including guidance as shown in \Cref{fig:teaser} $b.$.

The general setting of having arbitrary continuous input-output parameters allows us to apply our tool to various optimisation and general problems~\cite{Kehrer2012VisualizationMultifacetedScientificData}. We specifically conducted an informal expert session for a metallurgy setting to gauge the applicability of our approach.
The novel tool, finally, provides \textbf{guided creation of cost functions} through interactive elements, \cgls{ucq}, \cgls{xai}, \textbf{explorable interpolation} towards optimal solution. 
The code is archived on Zenodo~\cite{Kantz2026ParamInter}.

\section{Related Works}
\label{sec:rel-works}

Prior tools and explorations have already tackled optimisation problems in conjunction with parameter space analysis, either from a multi-objective optimality perspective with many possible solutions (Pareto front), using parallel coordinates as the core visualisation paradigm, or by integrating \cgls{ucq} and \cgls{xai} into the decision-making process.

\textbf{Pareto Front:}
Works within the high-dimensional optimisation field focus on variation within the Pareto front and parallel coordinates~\cite{Chinchuluun2007SurveyMultiobjectiveOptimization, Goguelin2017DashboardAdditive, Cajot2019InteractiveOptimizationParallelCoordinates, Cibulski2020ParetoFrontVisualization,Guo2011NuggetBrowser} to visualise multiple solutions jointly. Prior work to apply these principles within visualisation explore joint cost-optimisation problems in various settings~\cite{Chen2013SelfOrganizingPareto}. The tool enables the discovery of choices with the same costs, exploring the trade-offs between various cost-optimal points. Further, works also investigate overlaying markers for the underlying dimensionality-reduced space to explore their distributions~\cite{Raval2024Hypertrix}. Existing visual analytic tools, furthermore, already use parallel coordinates to display Pareto-optimal solutions, visualising them in an integrated fashion~\cite{Cajot2019InteractiveOptimizationParallelCoordinates,Cibulski2020ParetoFrontVisualization}.

\textbf{\cgls{xai} and Uncertainty visualisation:}
Integrating \cgls{ucq} into visualisation has been shown to be advantageous~\cite{Marusich2024UsingAIUncertaintyQuantification, Bhatt2021UCTransparency, Cresswell2024ConformalPredictionSetsImproveHumanDecisionMaking}, with certain visual parameters being more effective compared to others -- especially full \glspl{pdf} over \glspl{ci}~\cite{Sarma2024OddsInsightsUCQ}. All cited works, nevertheless, make the case that integrating uncertainty into user interfaces often improve the quality of decision-making and user trust. Additionally, \cgls{xai} is already used in some applications to foster transparency within tools, while there are still open challenges like deployment and evidence for long term benefit within applications~\cite{Garn2025TransparencyCombinatorialOptimisations,Kucher2025VisualAnalyticsExplainableAIIndustrialApplications}

However, most prior works focus on specific, singular solutions with no sensitivity analysis, hindering the comparison of instances. Hence, we introduce a novel \textbf{interpolation scheme} to search for optimal data points across the \textbf{input and output spaces}, supported by \cgls{ucq} and \cgls{xai} to guide the user.

\section{Interpolation through the Parameters}
\label{sec:meth}
\newcommand{\spanx}{\ensuremath{\bar{\x}\in\X, \bar{\x}=\{\bar{x}^0, \ldots, \bar{x}^k\}}}
Our interpolation approach within \toolname smoothly transitions from a starting point chosen from the input parameters, $\x_0\in\X$, to an optimal point, $\x_1\in\X$. We start from a selection
$$\x_{\mathrm{sel}} \in \left\{(x^0, \ldots, x^l)^T \,\Big|\,\min_{\spanx}\bar{x}^i \leq x^i\leq \max_{\spanx}\bar{x}^i\right\}\,.$$
The selected input parameters are used to search for the $k$ closest samples $\x_{\mathrm{knn}}\in \X$ from the provided input dataset $\x\in\X$ using \cgls{knn}, from which the start sample $\z_0 \in \Z$ is chosen, with
$\x=(z^0,\ldots,z^l)^T$
and
$\y=(z^{l+1},\ldots,z^{l+1+j})^T$, i.e. $\z\in\Z$ is the combination of input and output parameters. This combination allows us to include both input and output variables of a system into our cost and filter functions. $l$ and $j$ are the number of input and output parameters, respectively.
The other end of the interpolation, i.e. $\z_1$, is determined by a user-specified composite cost function of the form $o(\z)=\norm{\mathbf{d}(\z)}$ with $$ \mathbf{d}(\z)=(z-z^{*,h},\ldots\;|\; h\in H)^T\,, $$ with $H$ being the set of chosen target parameters indices, and $z^{*,h}$ being the chosen target value corresponding to the index. The minimum of this cost function is chosen as the target for interpolation by evaluating the cost $o(\z)$ for each value within the dataset $\Z$ after applying a user-specific filter operation on the data set $\sZ=s(\Z)$. Our tool supports bounding each variable for now, as seen in the next section. This leads to the optimum $\z_1=\arg\min_{\z\in\sZ} o(\z)$.

Once the final data sample $\z_1$ has been chosen, we interpolate between them in both input and output space. First, the input space is interpolated using the convex combination $$\x_{\lambda}^*=\lambda \x_0 + (1-\lambda) \x_1 \,,$$ over a discretized $\lambda=(0,1)$. We then feed each sample into a learned model \cgls{lgbm} ensemble $\bm{f}^*$~\cite{Zhang2017LGBM} to get all outputs $\y^*_{\lambda} =\bm{f}^*(\x_{\lambda}^*)$, and stacking $\x^*$ and $\y^*$ into $\z^*$. These samples are then used to find the closest real data point within the filtered set with a composite cost, i.e. $$\z_{\lambda}=\arg\min_{\z\in\sZ} \norm{\z-\z_{\lambda}^*}+\abs{o(\z)-o(\z_{\lambda}^*)}.$$

Only the resulting nearest data point is then used for visualisiations and interpolation, keeping the shown data within the set.

\section{Visual Representations}

The introduced interpolation principle is incorporated into the proposed \emph{visual} analytics flow to yield our \toolname. We will show the progression by introducing our visual elements one-by one.

\begin{figure*}[bt]
    \centering
    \begin{tikzpicture}[
            lens_txt/.style={
                    rectangle, draw=OliveDrab, fill=OliveDrab!10, text width=3cm, align=center, line width=2pt
                },
            lens_target/.style={
                    rectangle, draw=OliveDrab, fill=OliveDrab!10,  minimum size=0.8em, anchor=center, line width=2pt
                },
            lens_path/.style={draw=OliveDrab!80, line width=1pt,line cap=round, Round Cap-Round Cap, shorten <=-2pt, shorten >=-2pt},
            t1/.style={draw=LightCoral, fill=LightCoral!10},
            t2/.style={draw=Moccasin, fill=Moccasin!10},
            t3/.style={draw=SkyBlue, fill=tsnebg},
            t4/.style={draw=SkyBlue, fill=tsnebg},
        ]
        \node (main_img)[]{\includegraphics[width=0.45\linewidth]{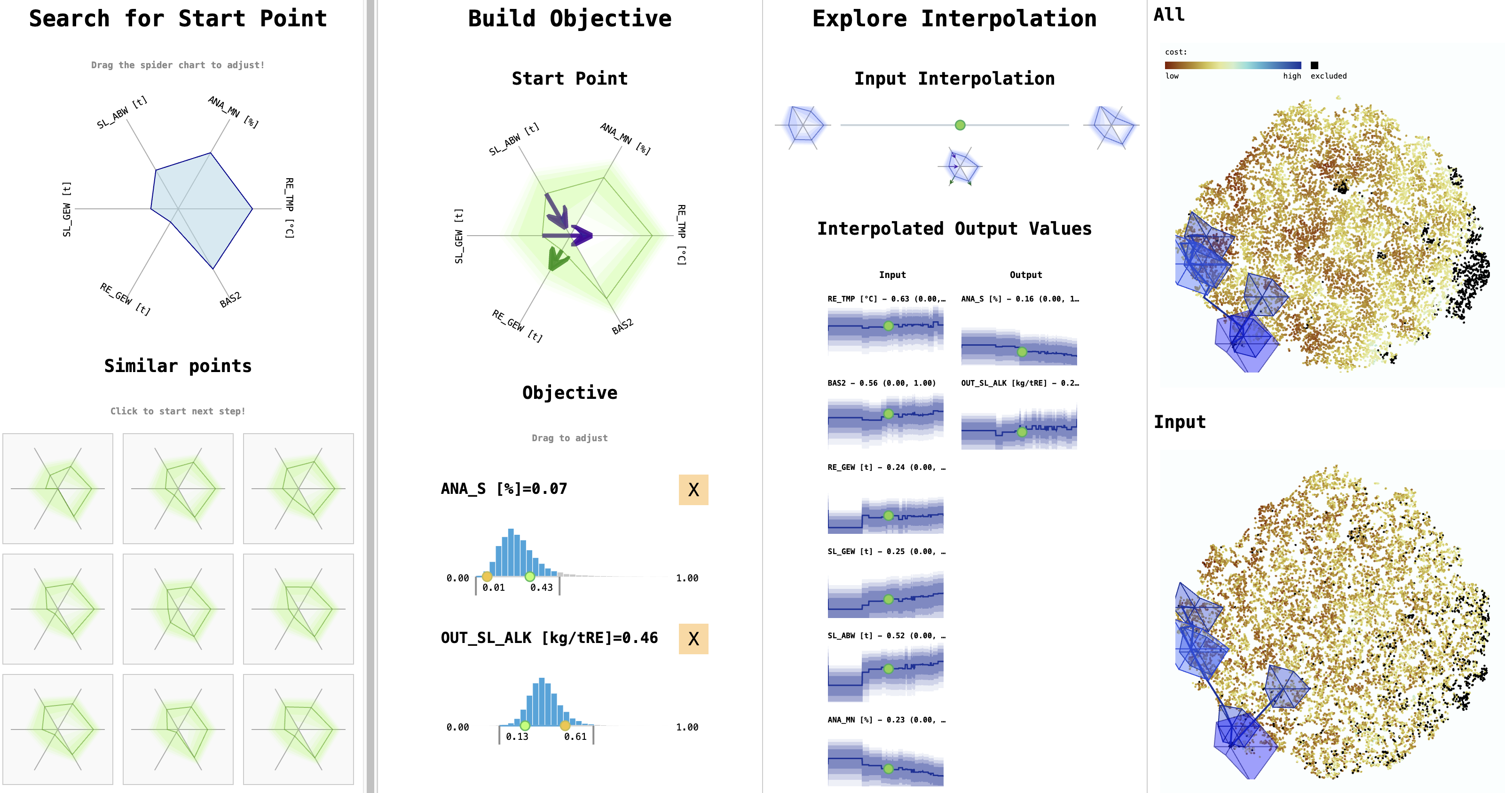}};
        \node (l_t_1)[lens_txt, t1, above left =1.25cm and 0.5cm of main_img.west] {$(a)$ Selecting the start point $\z_0$};
        \node (l_t_2)[lens_txt, t2, below left =1cm and 0.5cm of main_img.west] {$(b)$ Building the optimiser function $o(\z)$};
        \node (l_t_3)[lens_txt, t3, above right=1.25cm and 0.5cm of main_img.east] {$(c.1)$ Detailed interpolation view of $\z_\lambda$};
        \node (l_t_4)[lens_txt, t4, below right=1cm and 0.5cm of main_img.east] {$(c.2)$ \cgls{tsne} overview};

        \node (l_ta_1)   [below right= +2.5cm and +0.5cm of main_img.north west,lens_target, t1]{};
        \node (l_ta_2)   [below right= +0.2cm and +3.3cm of main_img.west,lens_target, t2]{};
        \node (l_ta_3)   [below right= +1.5cm and +1.5cm of main_img.north,lens_target, t3]{};
        \node (l_ta_4)   [below left = +0.2cm and +1.3cm of main_img.east,lens_target, t4]{};

        \path[lens_path, t1] (l_t_1.south east) -- (l_ta_1.north west);
        \path[lens_path, t2] (l_t_2.north east) -- (l_ta_2.south west);
        \path[lens_path, t3] (l_t_3.south west) -- (l_ta_3.north east);
        \path[lens_path, t4] (l_t_4.north west) -- (l_ta_4.south east);
    \end{tikzpicture}
    \caption{Overview over the full interface of  \toolname. $(a),(b),(c)$ show the stages of the exploration flow within for the loaded dataset and use case; the blast furnace optimisation. We note that the dataset is normalized to $[0,1]$. The expert selects the start using the radar chart as the input method in $(a)$, then sets up the cost function $o(\z)$ and filter $s(\Z)$ in $(b)$, and can finally observe the interpolation over the values in $(c.1)$ and across the \cgls{tsne} in $(c.2)$.}
    \label{fig:overview_interface}
\end{figure*}

\newcommand{\sideimage}[1]{
    \begin{wrapfigure}{l}{0.09\linewidth}
        \vspace*{-1em}
        \includegraphics[width=0.08\textwidth]{#1}
        \vspace*{-2em}
    \end{wrapfigure}
    \noindent
}
\sideimage{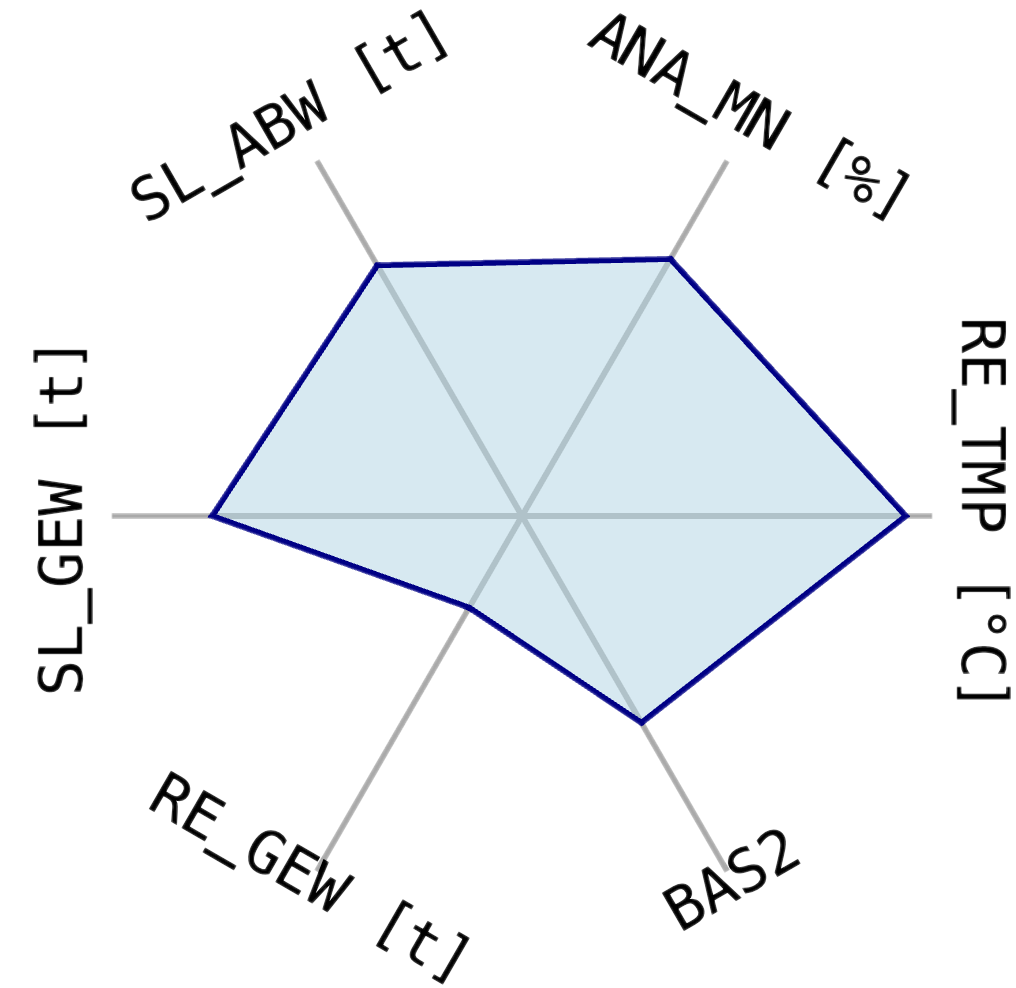}
The selection process for $\x_{\mathrm{sel}}$ is driven by a user-editable radar chart, as shown in the small glyph to the left, where the rough input parameter settings can be configured, with the $k$ nearest possible actual data points as selectable choices for starting points $\z_0$. We use radar charts as the central glyph throughout our interface, as they allow easy editing within the space they require, despite their flaws~\cite{Albo2015OffTheRadar}. The radar charts are also used, whenever appropriate, to convey uncertainty within the data, as we display it directly within the retrieved chart, allowing direct comparison between dimensions and harnessing the glyph space more effectively.

\sideimage{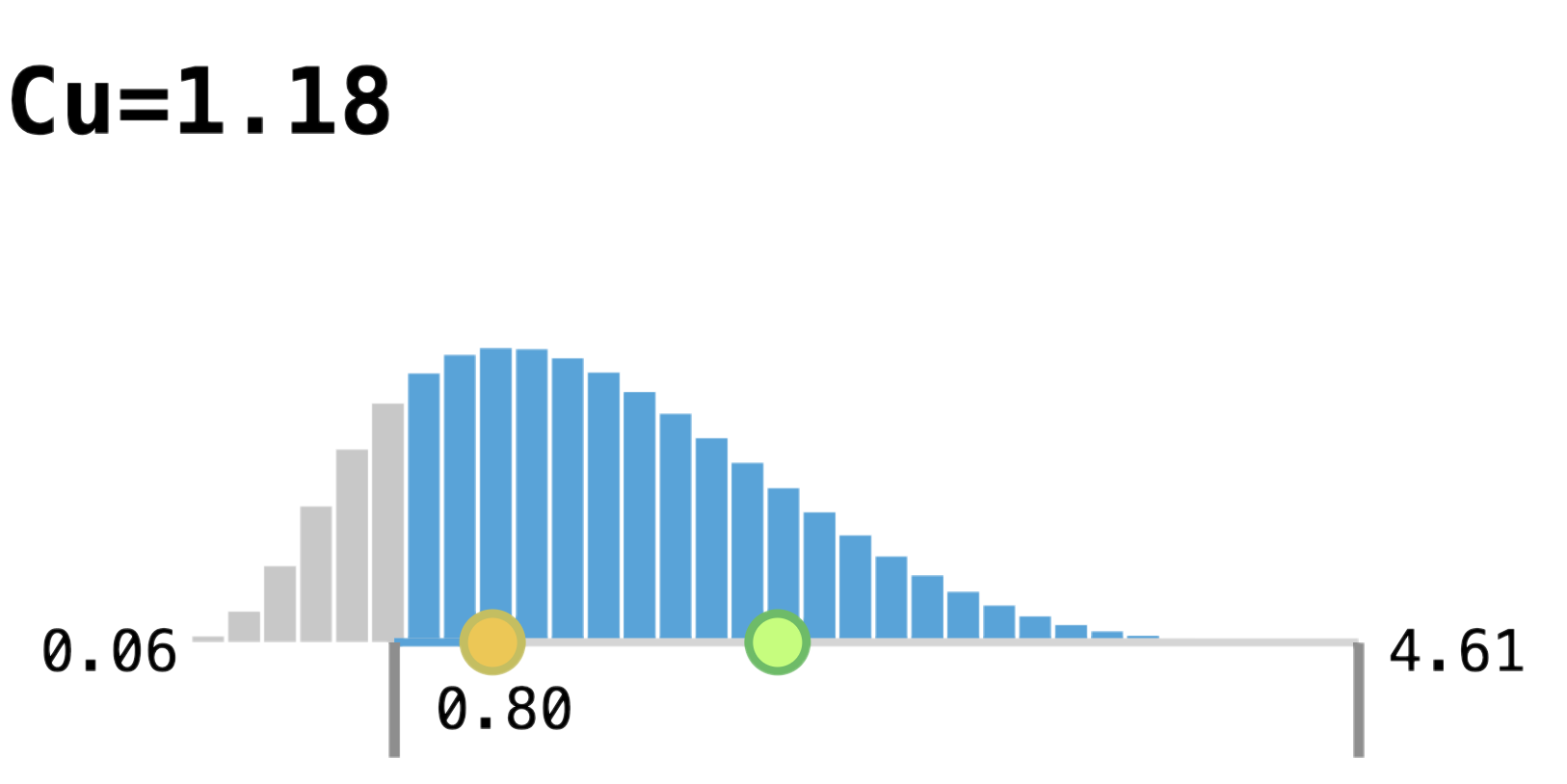}
The optimisation function $o(\z)$ can be tuned by adjusting variables from the input and output columns using an interactable histogram shown on the left, once the initial input parameter set has been chosen. Selecting a parameter adds it to the set $H$, which allows the user, through the histogram tool, to both define the target points $z^{*,h}$ by dragging the target point in yellow, and crop the data region for later interpolation, effectively defining $s(\Z)$.


\sideimage{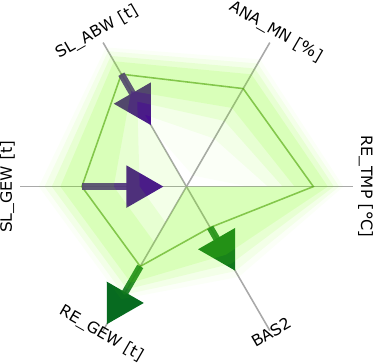}
We also display a sensitivity indicator on the radar chart whenever the user hovers over an output parameter. This indicator is based on a learned \cgls{lgbm}~\cite{Zhang2017LGBM} ensemble on the individual parameter $y^i$, i.e. $y^i=f^{*,i}(\x)\,,$ $i$ being the output dimension. We calculate the sensitivity $\bm{w}^{i}=\Phi_{f^{*,i}}(\x)$ using the \cgls{xai} method \cgls{sg}, effectively approximating the gradient $\nabla_{\x}f^{*i}(\x)$. This method has been shown to be quite robust in the face of noise, especially when combined with a tree-based \cgls{ml} method and densely sampled input space. It has also been tested in an applied industrial setting, similar to our use cases demonstrated in this paper~\cite{Kantz2024Robustness}. The adjustments of the output parameters is used to provide the users an estimate for the interpolation calculation by providing data-driven effect estimation. The uncertainty estimates are also shown as shaded areas, providing an estimate of the data variability at this point. We use an uncertainty-aware \cgls{vae} with a \cgls{gp} as the decoder~\cite{Tran2023FullyBayesianAutoencoders}. Prior research has shown that a complete view of the \cgls{pdf} is advantageous in single-dimensional settings for more accurate decision-making~\cite{Sarma2024OddsInsightsUCQ}.


\sideimage{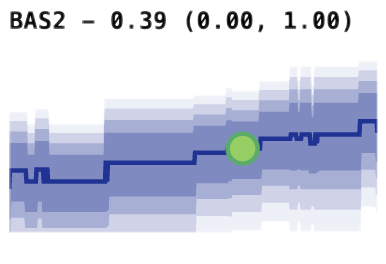}
The changes in each output variable $y_\lambda^{*i}$ relative to the interpolated variables are, furthermore, shown as small multiples, enriched with uncertainty estimates for all dimensions, complementing the visualisations within the radar charts. Users can hover over the full range to get the value for the data point at that specific progression, enabling them to \enquote{scrub} through the interpolation, related to how videos are usually scrubbed through~\cite{Schoeffmann2015VideoInteractionTools}.

\sideimage{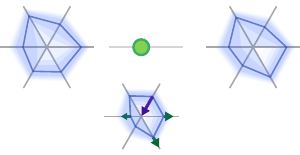}
\toolname, furthermore, provides a view of the interpolation between the input parameters, with explanations and uncertainties shown, again on top of the radar charts and updated as the user scrubs through the interpolation. This interface is linked to all small charts and shows the values for the specific data point simultaneously during scrubbing.

\sideimage{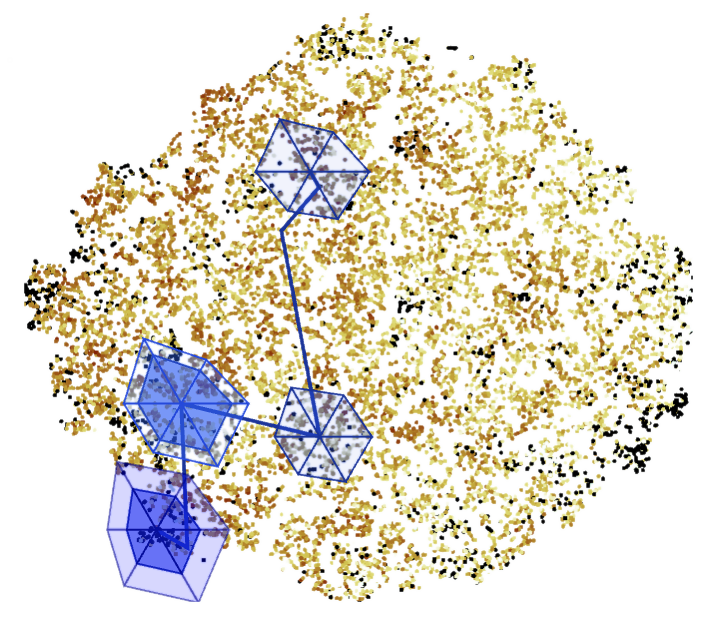}
Finally, the interpolations are overlaid on a \cgls{tsne}-reduced~\cite{vanderMaaten2008tSNE} scatter plot of the various parameter groups, e.g. input, output, and combined using fast GPU-accelerated libraries~\cite{Raschka2020cuml}. These provide users with an indication of the extent of the interpolation across the entire data space. We chose a nonlinear interpolation to reveal possible hidden close relations in the data, which a linear embedding like \cgls{pca}~\cite{Mackiewicz1993PCA} can not. Any linear projection also produces simple linear paths for our interpolation scheme, i.e. a linearly projected $\hat{\z}_{\lambda}$ will stay linear as
\begin{align*}
    \hat{\z}_{\lambda} & = \mathbf{P}\z_{\lambda} = \mathbf{P} (\lambda \z_0 + (1-\lambda) \z_1)  =\lambda\mathbf{P}   \z_0  + (1-\lambda) \mathbf{P}  \z_1\ = \\
                       & = \lambda \hat{\z}_{0} + (1-\lambda) \hat{\z}_{1}\,.
\end{align*}
Much more interesting insights of the high-dimensional space can therefore be achieved using non-linear layouts, which has been shown to be advantageous when grouping and imposing structure~\cite{Eckelt2022StructureEmbeddings}. This grouping, however reduces the semantic meaning of our interpolation lines, which has to kept in mind during analysis~\cite{Cashman2025CriticalAnalysisDR}. The \cgls{tsne} plot is enriched by a linked display of both the cost function $o(\z)$ and filter $s(\z)$ applied to each data point, while compositing the function. 
We use color mappings~\cite{Crameri2020ColorCommunication} to present costs in an accessible, contrast-rich view, even in dense or noisy regions. Users can still refer to the scrubbing view to get actual data values for their exploration.

This flow enables the users to iteratively refine their parameters, alternating between input and output spaces, while being alerted to sensitivities with respect to the outputs. \toolname, therefore fosters the discovery of advantageous parameters by showing the intermediate steps between a target output sample and initial input parameters, while also providing iterative means to exploration if the parameter set is not yet satisfactory through small multiples and guidance by low-dimensional representation. The user is, however, also free to choose any points in the low-dimensional space by hovering over them to reveal the input parameters and set them as a new start point.

\section{Case Study: Blast Furnace}
\label{sec:results}

We demonstrate \toolname on a use case co-developed with experts from the steelmaking industry, a blast furnace process optimisation.
Our specific use case concerns the output of bases via the slag to prevent instabilities in the blast furnace~\cite{Geerdes_2020BlastFurnaceIronmaking}, while also keeping sulfur output via the hot metal minimal advantageous for downstream processes. The problem has six input variables comprising different parameters for the blast furnace operations, i.e. temperature, slag and iron output, manganese concentration, and basicity. These parameters need to be adjusted for these two output parameters. We gathered about $16,000$ samples for this analysis. The specific optimisation problem was raised by industry experts and requires the discovery of parameters using an informed approach based on a repository of prior process data. We conducted an open-ended expert interview series, where we first demonstrated the tool, then gathered feedback. 
These were conducted with our two co-authors from the voestalpine Stahl GmbH. 
Their feedback was then incorporated into the tool, presented again with the updates; and finally provided to the experts for their down stream use. The tool is now deployed for their own internal use. We want to outline the analysis process the experts would use to come to a decision for their parameters next.

When applying \toolname to this data to arrive at decisions about operating procedures-- as in \Cref{fig:overview_interface} -- the metallurgy experts can start by selecting the appropriate or current start point $\z_0$ of the blast furnace, as seen in part $(a)$. This adjustment is performed by dragging the values in the radar chart to the desired position and then selecting one of the proposed data points.
Next, the experts can adapt the optimiser function $o(\z)$ within $(b)$ by interacting with the histograms. The expert chooses the sulfur output (\texttt{ANA\_S})
and base output (\texttt{OUT\_SL\_ALK}), then clips away the extreme tails of the histogram as they represent extreme operating regions not usually visited. Finally, the target values are set using the circle within the histogram, setting everything up for the interpolation.
The \cgls{tsne} output in $(c.2)$ is already updated, indicating that specific regions are omitted and showing the most advantageous areas right away. The radar charts additionally reveal, through their change towards a bigger area, that the parameters will most likely increase. While the expert is aware of the influence of the process parameters on each of the targets individually, it is a delicate task to control the blast furnace process in a way that both are optimal simultaneously. The full effect can only be understood by investigating the interpolation in $(c.1)$ -- it directly shows in which direction the parameters have to move, once the interpolation progresses towards the target -- both using \gls{xai} and the line charts containing the uncertainties for each interpolated point. The case at hand requires a rise in almost all input parameters to increase the base content and decrease the sulfur content in the output, as already indicated in the \cgls{tsne} overview. The specifics of the progression can be investigated as well, as the expert can hover over a specific part of the interpolation to retrieve the relevant output value, while also getting a visual indication of the distributional uncertainty at that point.

The interpolated radar chart at the top provides additional intuition into the sensitivity of the parameters, offering further insights into which direction might be advantageous to optimise in the next step. These insights would therefore allow the experts to tune the process parameters according to the final input variables, or use another operating point on the path to the optimum that might be less optimal, but offer a lower uncertainty, i.e. display narrower bands in the interpolation view, or be more stable with regard to the outputs, which can be identified through the \cgls{xai} markers.
The experts also remarked on some limitations of the interface, especially in regard to the radar plots -- as the precise configuration of inputs is difficult using the current interface, and would require improved facilities.



\section{Discussion \& Conclusion}
\label{sec:disc}

Our tool, \toolname, provides an iterative, visual, and guided optimisation tool that enables experts to discover input-output relations of parameters and achieve ideal compositions through interpolation. The optimisation process uses guidance to select input parameters, build an optimisation function, and interpolate towards the optimum. \toolname is already deployed for the blast-furnace use case, strengthening our case for real-world applicability.

Improvements to the visual guidance could include integration of the sensitivity results to steer the user not only visually but also to incorporate gradient estimation into the interpolation algorithm. 
Further research could also examine the decision-making capabilities of the \cgls{ucq} and \cgls{xai} integrations, especially in our industrial use case.
\section*{Acknowledgements}
This work is partially supported by the HEREDITARY Project, as part of the European Union's Horizon Europe research and innovation programme under grant agreement No GA 101137074.
\printbibliography

\appendix
\section{Use cases}
We also provide further use cases to illustrate the versatility of our tool.
\subsection{Alloy Mixture Optimisation}
\begin{figure}[tb]
    \centering
    \includegraphics[width=0.6\linewidth]{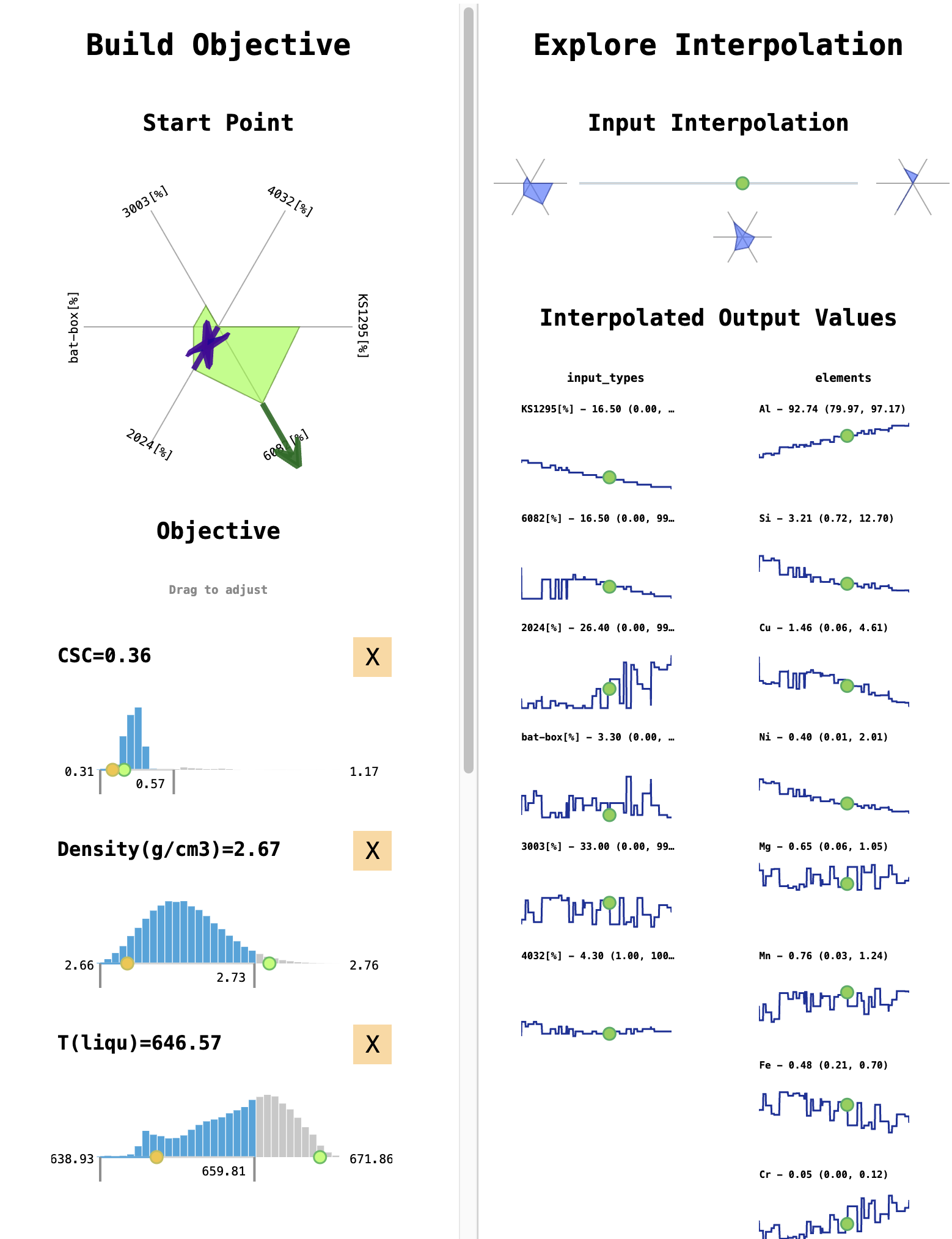}
    \caption{Setting up the targets for the 2025 SciVis contest dataset on the left, and observing the interpolations on the right panel. The goals for a low crack susceptibility, low density, and low melting point are set up using the interface, and the interpolation towards the optimum is shown.
    }
    \label{fig:scivis_usecase}
\end{figure}
Determining the composition of scrap metal to discover novel alloys is crucial to improving recycling rates. Enabling the ad-hoc discovery of viable compositions for additive manufacturing use cases of these materials was the goal of the 2025 SciVis Contest \cite{Bugelnig2024SciVisContest}. The challenge aims to support the discovery of optimal scrap-metal mixtures by promoting innovative approaches to guided visualisation of a simulated dataset. This dataset contains a sampled hypercube of the six input ratios and 64 resulting elemental compositions and various physical properties relevant for additive manufacturing, with $300,000$ samples available for analysis.

Metal-forming experts can also use \toolname for this use case. First, the expert can select a starting composition, then compose a target function tailored to their needs. The novel material might need to have low crack susceptibility while still melting at a low enough temperature to be melted within an additive manufacturing process. A third requirement for low density might be necessary to satisfy the requirement for a low-weight material. All three targets and the resulting interpolation are shown in \Cref{fig:scivis_usecase}, illustrating how this tool can be useful for this application by providing directions to optimise for, and suggesting local sensitivities using \cgls{xai}. The experts could use \toolname similarly to the blast furnace approach, with the mixture ratios requiring stability analysis.
We, however, disable the uncertainties view, as this simulated dataset has no inherent (aleatoric) uncertainties.


\subsection{Biomedical Parameter Discovery}

Determining the implications of parameter choices for simulations is key to the operation and starting of simulations, especially for biomedical applications. Models like these shine light on the biological processes and interactions within diseases, and with various approaches~\cite{Thomas2016CancerModeling,Norton2019CancerModeling,Cesaro2022Mast}. We apply our tool on the \cgls{mast} model~\cite{Cesaro2022Mast}, an integrated modeling approach for the tumor-immune system interaction. 
Our simulation dataset contains a grid search over a variety of simulation settings describing cell division rate, cure injection rates and the like. Simulating these yields spatially resolved cell grids, which we aggregate into abundance percentages. These are then treated as inputs in \toolname, allowing the experts to set targets of cell type abundances, yielding specific simulation parameters which they can them use for their next exploratory runs. Experts could use the ability to set the input parameters, in this case the tumor abundances to a low value, while also constraining the cure injection rate to estimate which other parameters are beneficial to lowering the tumor growth in the \cgls{mast} model.

This dataset and the SciVis contest set are loaded into our demonstrator and are available for exploration on \url{\demolink}.
\section{Implementation}
\label{sec:impl}

\toolname is implemented as a web application, using D3~\cite{Bostock2011D3} and its wrapper \verb|D3FC| (\href{https://d3fc.io/}{d3fc.io}) to build the visualisations. We use the WebGPU functionality of \verb|D3FC| to efficiently visualise the large sets of data points, and employ the \verb|cuml|~\cite{Raschka2020cuml} package in the backend to quickly compute dimensionality reductions. The code is available on \href{\codelink}{GitHub}.

\end{document}